\documentclass{aastex}          
\usepackage{spr-astr-addons}    

\newcommand{\zsun}{\ensuremath{\mathit{Z}_{\odot}}}                
\newcommand{\teff}{\ensuremath{\mathit{T}_{\rm eff}}}
\newcommand{\lsun}{\ensuremath{\mathit{L}_{\odot}}}

\begin{document}

\title{The theory of stellar winds}

\shorttitle{Stellar wind theory}
\shortauthors{Vink, J.S.}

\author{Jorick S. Vink\altaffilmark{1}} 

\altaffiltext{1}{Armagh Observatory, College Hill, Armagh BT61 9DG,
Northern Ireland, UK}

\begin{abstract}
We present a brief overview of the theory of stellar winds with a 
strong emphasis on the radiation-driven outflows from massive stars. 
The resulting implications for the evolution and fate of massive stars are also discussed. 
Furthermore, we relate the effects of mass loss to the angular momentum evolution, which is 
particularly relevant for the production of long and soft gamma-ray bursts. 
Mass-loss rates are not only a function of the metallicity, but are also found to depend on 
temperature, particularly
in the region of the bi-stability jump at 21 000 Kelvin. We highlight the role 
of the bi-stability jump for Luminous Blue Variable (LBV) stars, and discuss suggestions 
that LBVs might be direct progenitors of supernovae. 
We emphasize that radiation-driven wind studies rely heavily on the input opacity data and 
linelists, and that these are thus of fundamental importance
to both the mass-loss predictions themselves, as well as to our overall understanding of the 
lives and deaths of massive stars. 
\end{abstract}

\keywords{Mass loss, massive stars, metallicity, 
supernovae, luminous blue variables, gamma-ray bursts}

%

\section{Introduction}
\label{s:intro}

Despite their relative rarity, massive stars are dominantly dangerous
for their environments. This is a result of their mass and energy input via
stellar winds and subsequent core-collapse supernovae (SN). 
A overarching parameter for the life expectancy of a massive star concerns 
its strong mass outflow which is driven by radiative forces on millions of 
ionic spectral line transitions, providing it with its name 
``line-driven wind''. 

On the one hand, mass loss is thought to be a key agent in revealing 
chemically processed material at the stellar surface, making it responsible 
for evolutionary scenarios such as the 
O $\rightarrow$ Luminous Blue Variable (LBV) $\rightarrow$ Wolf-Rayet (WR) star 
$\rightarrow$ SN sequence 
(e.g. Conti 1976, Chiosi \& Maeder 1986, Langer et al. 1994).
Furthermore, it determines the stellar mass before collapse and is thus relevant 
for the type of compact remnant that is left behind (i.e. neutron star or 
black hole). 
On the other hand, the role of mass loss may be equally relevant for 
the loss of angular momentum (e.g. Meynet \& Maeder 2003). 

With respect to the latter, it has been suggested that low metallicity 
(actually low ``iron'' contents; Vink \& de Koter 2005) leads to less mass 
and angular momentum loss in low metallicity environments, perhaps resulting in a 
preference of long gamma-ray bursts (GRB) in the early Universe, but 
however interesting 
the metallicity dependence and the long GRB puzzle may be, the 
temperature dependence of stellar winds and its role in the 
angular momentum evolution of massive stars has been highlighted more 
recently with respect to the possibility of {\it bi-stability braking} 
(Vink et al. 2010). 

Given the crucial role that mass loss plays for massive star evolution,
we discuss the theory of massive star mass loss and its 
implications, with a focus on the metallicity ($Z$) and effective 
temperature ($T_{\rm eff}$) dependence. 
We will see that the {\it iron line opacity} plays a dominant role in both 
cases.   
 
\section{The theory of line-driven winds}
\label{s:theory}

The theory goes back to the early 1970s when Lucy \& Solomon (1970) 
suggested that selective radiation pressure on spectral lines 
is capable of driving stellar winds from the surface of massive stars, proving
an explanation for the P Cygni profiles observed in the ultraviolet spectra 
of O-type stars in the late 1960s. For an extensive overview of both the 
theory and recent observational developments, such as wind 
clumping, we refer the reader to Puls et al. (2008).
The theoretical framework is based around the momentum equation:

\begin{equation}
\label{eqn:mom}
v (\frac{dv}{dr})_{\rm Newton} = - \frac{G M_{\rm eff}}{r^2} +  g_{\rm line}
\end{equation}
where for simplicity the gas pressure has been ignored (as the radiative 
pressure is the much more dominant factor in most parts of the stellar 
wind).

An oft-cited paper concerns the study of Castor et al. (1975, hereafter CAK) who 
expressed the radiative line acceleration $g_{\rm line}$ 
as a function of the Sobolev velocity gradient ($g_{\rm line} = f(dv/dr)_{\rm Sobolev}$), 
involving a 2-parameter force multiplier ($k$,$\alpha$) parametrization.
Subsequent improvements such as the inclusion of improved line lists by the Munich group
led to a reasonably good agreement with observed values (e.g. Pauldrach 
et al. 1986). We note that the CAK-type wind dynamics
is based on the assumption that $(dv/dr)_{\rm Sobolev}$ plays the same role as  
$(dv/dr)_{\rm Newton}$, such that the critical point of the wind is no longer 
the sonic point (as in solar wind theory) but further downstream.
The validity of this approach has been questioned by Lucy (2007).
Furthermore, these works relied on the assumption that stellar photons 
could interact with the outflowing ions only once. 

In order to account for multiple 
scatterings Abbott \& Lucy (1985) developed a Monte Carlo methodology, which was updated 
and extended by Vink et al. (2000, 2001). In these oft-used mass-loss prescription the 
velocity law was originally adopted (based on a semi-empirical motivation), but 
this assumption has recently been alleviated by M\"uller \& Vink (2008) who express 
the line acceleration as a function of radius $g_{\rm line} = f(r)$ rather than the velocity
gradient as done explicitly in CAK theory. The implication is also that the critical point is 
mathematically the sonic point. 

The resulting mass-loss rates follow a scaling relation that only depends on the 
basic stellar parameters, approximately as: 

\begin{equation}
\label{eqn:mdot}
\dot{M} \propto Z^{0.7}  L^{2.2}  M^{-1.3}  (v_{\infty}/v_{\rm esc})^{-1.4}  T_{\rm eff}^{1.1}
\end{equation}
over the entire range of $T_{\rm eff} = 12.5 - 50$kK -- except for the
{\it bi-stability} jump around 25 kK, where the mass-loss properties are predicted to 
change drastically, with $v_{\infty}$ dropping by a factor of two (Pauldrach \& Puls 1990; 
Vink et al. 1999) 
and the mass-loss rate jumping upwards by a factor of five (Vink et al. 1999; 
see Sect.\ref{s:tdep}).

\section{The metallicity dependence and the long gammma-ray burst puzzle}
\label{s:zdep}

Massive stars rotate rapidly, with rotational velocities of up to  
400 km/s. This is understood to have dramatic consequences for their 
evolution and ultimate demise, which may involve the production 
of a long-duration gamma-ray burst (long GRB).

As a result of this rotation, the pole becomes hotter than the stellar 
equator (Von Zeipel theorem), which 
enables a rather complex meridional circulation  
in the stellar interior. During this process, nuclear 
processed material is transported from the core to the stellar 
envelope, thereby enriching the stellar surface with elements 
such as nitrogen (N), which is produced during the CNO cycle of 
hydrogen (H) burning.
Later on, the combination of mass loss and 
rotation also leads to the transfer of products of helium (He) burning to the 
surface, enriching the atmosphere with carbon (C) during the final 
carbon-rich Wolf-Rayet (WC) phases before the stellar core is expected 
to collapse, producing a supernova (SN) -- in some cases in 
conjunction with a long GRB.
 
\subsection{The collapsar model for gamma-ray bursts}

The most popular explanation for the long GRB phenomenon involves the 
collapsar model (MacFadyen \& Woosley 1999) in which a rapidly rotating core  
collapses and produces an accretion disk surrounding a black hole.
One of the persistent problems with the collapsar model was that 
the object not only requires a high rotational velocity at the 
very beginning of its life, but that it is required 
to maintain this rapid rotation until the end.
This is a significant challenge because 
strong stellar outflows are expected to remove 
angular momentum. Stellar models with rotation show that the objects not 
only remove up to 90\% of their initial mass in winds when they have reached 
their final Wolf-Rayet phase, but as a 
result of this wind, the stars are also expected to 
come to a complete standstill. 

The question is what is expected for Wolf-Rayet stars in low metallicity 
galaxies. Are they subject to similar mass and angular momentum loss? 

\begin{figure*}
\includegraphics[width=13cm]{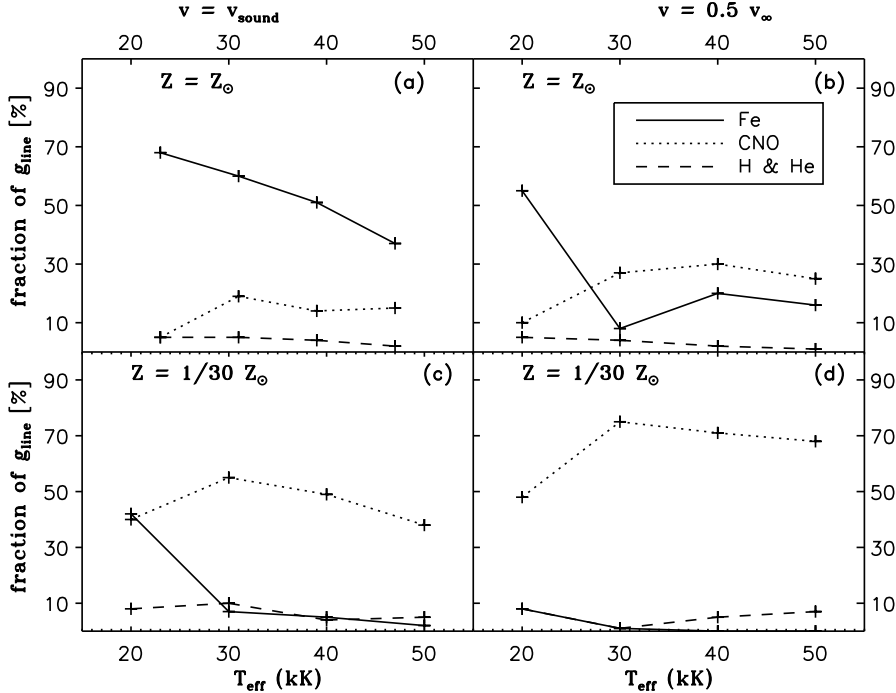}
\caption{The chemical make-up of the line acceleration for a range of different temperatures. The upper panels
(a and b) show the dominating species at solar metallicity. 
The lower panels (c and d) show the same for very low $Z$ (1/30 \zsun).
The left-hand panels depict the dominant species in the inner wind, where $\dot{M}$ is set, whilst the right-hand panels showcase
the chemical make-up in the supersonic portion of the wind that is responsible for the terminal velocity (from Vink et al. 2001).}
\label{fig:zrel}
\end{figure*}

\subsection{Iron iron iron}

Radiation hydrodynamic simulations show that stellar winds 
from massive O-type stars are driven by the radiation pressure 
on metal lines, and specifically on iron (Fe), despite the fact that it  
is such a rare element.
It should be noted that even in the metal ``rich'' environment of the 
Milky way the H abundance is already 2500 times larger than that of Fe.
However, owing to iron's highly complex atomic structure it has millions of 
line transitions, which makes it an extremely efficient absorber of radiation 
in the inner atmosphere around the sonic point, 
where the mass-loss rate is set (Vink et al. 1999, Puls et al. 2000).
Figure~\ref{fig:zrel} shows the contribution of the different chemical elements (H, He, CNO, F) to the 
total line acceleration for normal O-type stars. 
At solar $Z$, Fe dominates the line acceleration near the sonic point, whilst its contribution 
drops significantly at lower density. Therefore, the Fe line contribution
is less important in the supersonic region where the wind terminal velocity is set (panel b). 
And similarly at lower $Z$ (panels c and d). 
Here, the contribution of CNO and elements such as Cl, Ar, P and S becomes highly 
relevant instead (Vink et al. 1999, 2001).

Up to 2005 most stellar modellers assumed that due to the overwhelming presence of C in WC atmospheres, it 
would probably be C that drives Wolf-Rayet winds, rather than Fe.
This assumption also implied that WC stars in low metallicity ($Z$) 
galaxies would have stellar winds equally strong as those in the Galaxy, 
and still removing the required angular momentum.
It was for this reason that there was no satisfactory explanation 
for the long GRB puzzle. 
In 2005, we performed a pilot study of 
Wolf-Rayet mass loss as a function of $Z$ finding that 
although C may be the most abundant metallic element in WC atmospheres 
it is nonetheless the much more complex Fe element that drives the 
stellar wind (Vink \& de Koter 2005, Gr\"afener \& Hamann 2008). 
In other words, host galaxy 
metallicity plays a crucial role: objects that are born with fewer Fe 
atoms lose less matter by the time they reach the end of their lives, despite 
their larger contents of CNO material. The striking implication
is that objects in low metallicity
environments -- such as those characteristic for the early Universe -- can
 keep their angular momentum, enabling a potential GRB event. 
Interestingly, there indeed appears
to be a preference for long GRBs towards lower $Z$ systems (e.g. Vreeswijk et al. 2004).

\begin{figure*}
\includegraphics[width=10cm,angle=270]{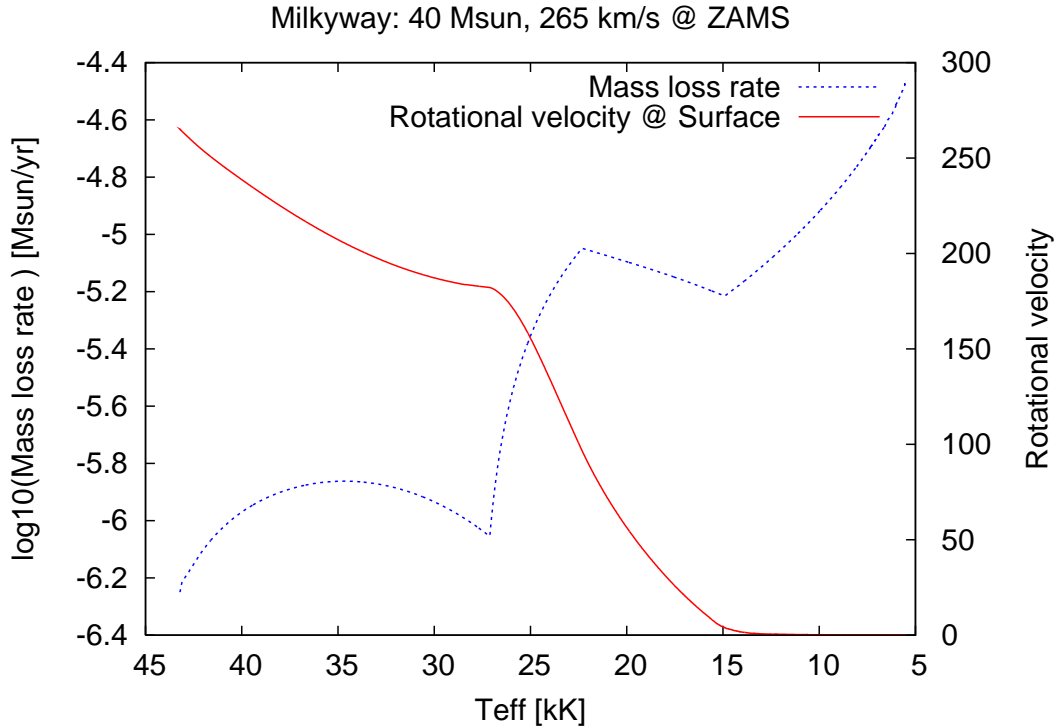}
\caption{The predicted mass-loss rate (dotted line) and rotational velocity (solid line) of a massive star
as a function of effective temperature (see Vink et al. (2010) and Brott et al. (2010) for 
further details.)} 
\label{fig:brott}
\end{figure*}

\section{The temperature dependence and the bi-stability jump}
\label{s:tdep}

As discussed at the end of Sect.~\ref{s:theory}, the mass-loss rates 
of massive objects are predicted to increase strongly when stars evolve towards
the B supergiant regime. This $\dot{M}$ jump is referred to as the bi-stability
jump. The mass-loss behaviour is depicted as the dotted line 
in Fig.~\ref{fig:brott}. Recent stellar models with rotation of Brott et al. (2010) 
that include this Vink et al. mass-loss jump show a dramatic braking (see solid line), which we
refer to as {\it bi-stability braking (BSB)}. BSB might explain the general
slow rotation of B supergiants. An alternative explanation for the slow rotation of 
B supergiants might involve a core He-burning nature for B supergiants (see Vink et al. 
2010 for a detailed discussion). 

The cause of the mass-loss bi-stability jump is that the most important line driving element Fe recombines 
from Fe {\sc iv} to Fe {\sc iii} at 25 000 K and that suddenly the Fe lines become 
much more effective as they fall in the wavelength range where the flux distribution is maximal. 
The result is an increase in $\dot{M}$ and a drop in terminal 
velocity. The latter has been confirmed in observed data-sets (e.g. Lamers et al. 1995), 
but the jump in mass-loss rate is still controversial (e.g. Crowther et al. 2006, Benaglia 2007, 
Markova \& Puls 2008).

The relevance for stellar evolution is that when massive stars evolve to 
lower \teff\ after the O star main sequence phase, they are expected to cross 
the bi-stability jump. Interestingly, LBVs brighter than log ($L/\lsun$) $= 5.8$ 
(see Fig.~\ref{fig:hrd}). 
are expected to encounter it continuously - on timescales of their photometric 
variability, which we discuss in the next section.

\section{Luminous Blue Variables and core-collapse supernovae}
\label{s:lbs}

\begin{figure}
\includegraphics[width=9cm]{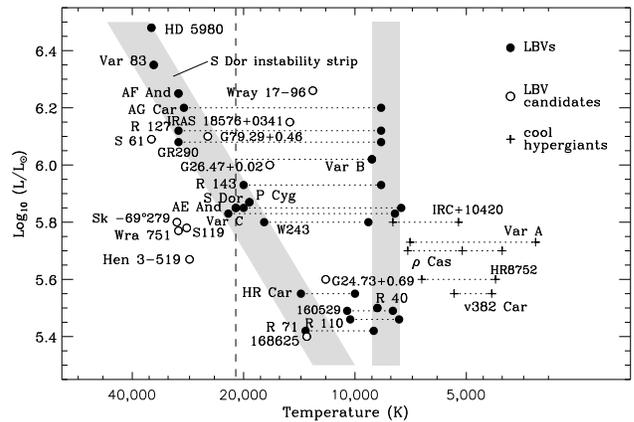}
\caption{The position of known local group LBVs in the Hertzsprung-Russell diagram. At visual minimum
LBVs are found along the slanted band as B supergiants, whilst they become F supergiants when observed 
at visual maximum (vertical band). 
Also indicated (dashed line) is the position of the mass-loss bi-stability jump at $\simeq$22 000 K 
(after Humphreys \& Davidson 1994; Smith et al. 2004; Vink 2009).} 
\label{fig:hrd}
\end{figure}

In the previous section, we discussed the physics of the bi-stability jump in the context of 
normal OB supergiants. The jump might also play a role in the mass-loss behaviour 
of LBVs, in particular with reference to the normal ``S Doradus'' type variations where the stars
change their effective temperatures from values as high as 30 kK (where they are identified as 
B supergiants) to approximately 10 kK (where they are identified as F supergiants). 
During these S Dor excursions, the stars keep crossing the 
temperature range of the bi-stability jump (see Fig.\ref{fig:hrd}), thereby likely inducing 
variable mass loss.

Such variable mass loss is in turn understood to be responsible for a non-uniform 
circumstellar medium, which could show up in the lightcurves 
and spectra of core-collapse SNe -- if LBVs were in an advanced enough evolutionary state. 
Current wisdom is that LBVs are not evolved enough (e.g. Langer et al. 1994) and that the LBV
phase of evolution is the stage in which most of the outer hydrogen envelope is lost, prior  
to the objects turning to the much hotter WR phase. During this stage, 
they are thought to be subject to core He-burning for another couple of 100 0000 years before 
they finally run our of fuel and collapse. 

However, there have been recent observational hints that LBVs might explode early. 
Kotak \& Vink (2006) noted that the variable radio emission of some transitional core-collapse 
SNe such as SNe 2001ig and 2003bg might be the result of variable LBV progenitor mass loss, which 
was subsequently followed up with spectroscopic wind velocity variations in SN 2005gj (Trundle et al. 
2008). In the meantime, Smith et al. (2007) have argued for 
LBV super-outburst circumstellar material being responsible for a number of 
luminous interacting SNe, whilst Gal-Yam et al. (2007) 
identified an LBV-type hypergiant star in the pre-explosion image of SN 2005gl. 

In other words, there appears to be a growing body of observational evidence that 
LBVs might explode already during their LBV phase of evolution. It might turn out to be 
an interesting puzzle to reconcile these findings with the theory of stellar structure and 
evolution.

\section{Final words}
\label{s:sum}

In this communication, we have presented a couple of 
examples of the type of information that mass-loss predictions can provide for our
understanding of the evolution and fate of massive stars, especially with respect 
to SNe and long GRBs. 
The manner in which these mass-loss predictions 
depend on the basic stellar parameters, such as $Z$ and $T$, can ultimately only be 
as good as the underlying atomic data. 
Extensive opacity data and linelists, such as those provided 
by Kurucz \& Bell (1995) that have been used in our study, are thus of fundamental importance
not only for the mass-loss predictions themselves, but also for understanding  
the lives and deaths of massive stars -- over all cosmological epochs.

%


%

%

\end{document}